\documentclass[12pt,notoc]{JHEP3}
\usepackage[fleqn]{amsmath}
\usepackage[latin1]{inputenc}
\usepackage{amssymb,bbm}
\usepackage{boxedminipage}

\def\del{\partial}
\def\e{\mathrm{e}}

\def\sign{\mathrm{sign}}
\def\ket#1{\big|#1\big\rangle}
\renewcommand*{\i}{\mathrm{i}}
\newcommand*{\ie}{i.\,e.\ }
\newcommand*{\eg}{e.\,g.\ }

\renewcommand*{\d}{\mathrm{d}}
\newcommand*{\dx}{\d x}

\newcommand*{\ds}{\d s}

\def\bb#1{\mathord{\hbox{\boldmath$#1$}}}
\def\N{{\mathbbm N}}
\def\Z{{\mathbbm Z}}
\def\id{{\mathbbm 1}}
\def\calI{{\cal I}}
\def\calJ{{\cal J}}
\def\calK{{\cal K}}
\def\calP{{\cal P}}

\title{
Type IIB tensionless superstrings\\ in a pp-wave background
}

\author{
  Andreas Bredthauer\footnotemark[1], 
  Ulf Lindstr{\"o}m\footnotemark[1],\footnotemark[4] 
  Jonas Persson\footnotemark[1], 
  Linus Wulff\,\footnotemark[3]\\
  \footnotemark[1]\,\,\parbox[t]{13cm}{
    Department of Theoretical Physics, Uppsala University\\
    Box 803, SE-751\,08 Uppsala, Sweden
  }\\ \ \\ 
  \footnotemark[4]\,\,\parbox[t]{13cm}{
    Helsinki Institute of Physics\\
    P.O. Box 64, FIN-000\,14 University of Helsinki, Finland
  }\\ \ \\
  \footnotemark[3]\,\,\parbox[t]{13cm}{
    Field and Particle Group, Department of Physics, Stockholm University\\
    AlbaNova University Center, SE-106\,91 Stockholm, Sweden
  }  
  \ \\ \ \\  
  {\tt Andreas.Bredthauer, Ulf.Lindstrom, Jonas.Persson@teorfys.uu.se,\\
   \tt Linus.Wulff@physto.se}
}
\abstract{
We solve the tensionless string in a constant plane wave background and obtain
a hugely degenerate spectrum. This is the case for a large class of plane wave
backgrounds. We show that the solution can also be derived as a consistent
limit of the quantized tensile theory of IIB strings in a pp-wave. This is
in contrast to the situation for several other backgrounds. 
}

\preprint{
  hep-th/0401159\\
  UUITP-01/04\\
  USITP-04-01\\
  HIP-2004-02/TH
}

\begin{document}
\section{Introduction}\label{sec:intro}
Although the study of tensionless strings has a long history
\cite{Schild:vq}-\cite{Zheltukhin:2001jw}, it has become of more wide-spread
interest only recently. Originally conceived as a limit of the same type as
that of massless relativistic particles and of similar interest for the study
of high-energy behavior, it is presently perhaps studied mainly for its possible
relations to higher spin theory 
\cite{Sezgin:2002rt,Sagnotti:2003qa,Lindstrom:2003mg}.

Interest in the relation to higher spin theory was triggered by the observation
that such a spectrum seems to arise on the Yang-Mills side in the AdS/CFT
correspondence \cite{Haggi-Mani:2000ru,Sundborg:2000wp}, where the limit 
$T\to 0$ corresponds to free Yang-Mills. To really corroborate the
correspondence in this limit, one should display the spectrum of tensionless
strings in the bulk, \ie\ in AdS space. However, quantizing the classical
tensionless string in that space meets with similar problems as does quantizing
the tensile string \cite{Parnachev:2002kk,Callan:2003xr}.\footnote{See, 
however, \cite{Bonelli:2003zu}.} A more modest ambition which we settle for
here is to study the spectrum of tensionless strings in a Penrose limit of
$AdS_5\times S^5$, \ie\ in a maximally supersymmetric pp-wave background
\cite{Blau:2001ne,Blau:2002dy}.

In this paper we first quantize the classical tensionless (super)string in the
relevant pp-wave background and study its spectrum. We find a hugely
degenerate spectrum which may be classified in terms of affine Lie-algebras. 
Then, we move on to study the quantized tensile (super)string first discussed
by Metsaev and Tseytlin \cite{Metsaev:2001bj,Metsaev:2002re},\footnote{A good
introductory review can be found in \cite{Plefka:2003nb}.}  recovering the
same spectrum (and the same model) in a certain well-defined limit.  

Apart from shedding light on the behavior of quantized tensionless strings in
this kind of curved background, our results also illuminate the long-standing
question of whether the tensionless limit commutes with quantization: In this
case it does. At the same it has been shown that this is not the case, \eg
for strings moving in backgrounds that are only approximate string solutions
\cite{Iannis}. We would argue that the reason quantization commutes with the
tensionless limit in our case is that the pp-wave background we consider is a
solution to all order in $\alpha^\prime$. This argument might seem to
contradict the difficulties in showing commutation in a flat background. An
important difference between these two cases, however, is the existence of a 
mass scale in the pp-wave to which the tension may be compared. It is also
this mass scale which seems to allow a straight-forward generalization of our
results to homogeneous plane waves. These are backgrounds slightly more
complicated than the one we study and have recently been discussed
\cite{Blau:2002js,Blau:2003rt}.

Finally, we note that the tension and mass scale always appear in such a 
combination that the tensionless limit is equivalent to the mass becoming
infinitely large, corresponding to an infinite background curvature. 

The paper is organized as follows: In section 2, we derive the action for
the tensionless IIB string in the (maximally supersymmetric) constant plane
wave background starting from the corresponding results in the tensile case.
Section 3 contains the classical and quantized solutions to this model and
derive the spectrum.  In section 4, we obtain our previous results as a
well-defined limit of the tensile string solutions in this background. Section
5 deals with the homogeneous plane waves. The paper is concluded with a short
discussion and outlook in section 6. 

\section{The action for a tensionless superstring in a pp-wave background}
\label{sec:action}
In this section, we derive the tensionless superstring action in the
constant plane wave background. This is a special plane wave with a constant
Ramond-Ramond \hbox{5-form} field strength given by\footnote{$I=1...8$ denote
the eight transverse coordinates while $x^\pm=\frac{1}{\sqrt{2}}(x^9\pm x^0)$
are the usual light-cone coordinates.} 
\begin{eqnarray}
  \ds^2 = 2 \dx^+ \dx^- - \mu^2 x^I x_I \dx^+\dx^+ + \dx^I\dx_I,
  \hspace{0.5cm}F_{+1234}=F_{+5678}=2\mu.
  \label{eqn:metric}
\end{eqnarray}
The type IIB superstring in this background was first discussed by Metsaev
\cite{Metsaev:2001bj} and then solved by Metsaev and Tseytlin
\cite{Metsaev:2002re}. For simplicity, our notation closely follows that of the
latter paper.  A short overview can be found in appendix \ref{app:notation}.
Below, we derive the corresponding light-cone Lagrangian for a tensionless
superstring starting from the $\kappa$-symmetry fixed Lagrangian given in
\cite{Metsaev:2001bj}:
\begin{eqnarray}
  \mathcal{L}=-\frac{T}{2}\sqrt{-g}g^{ab}h_{ab}
  +\i T\epsilon^{ab}\partial_{a}X^{+}
   \left(\theta^{1}\bar{\gamma}^{-}\partial_{b}\theta^{1}
  -\theta^{2}\bar{\gamma}^{-}\partial_{b}\theta^{2}\right). 
  \label{eqn:ksymfixedL}
\end{eqnarray}
Here, $T$ is the string tension and
\begin{eqnarray}
  h_{ab} & = & 2\partial_{a}X^{+}\left(\partial_{b}X^{-}
    +\i\theta^{1}\bar{\gamma}^{-}\partial_{b}\theta^{1}
    +\i\theta^{2}\bar{\gamma}^{-}\partial_{b}\theta^{2}\right)\nonumber\\
    &&-\left(\mu^{2}X^{I}X_{I}
    +4\i\mu \theta^{1}\bar{\gamma}^{-}\Pi\theta^{2}\right)\partial_{a}X^{+}
    \partial_{b}X^{+}
    +\partial_{a}X^{I}\partial_{b}X_{I}.
\end{eqnarray}
To be able to take the tension $T$ to zero, we go to the phase space Lagrangian
in the usual way \cite{Lindstrom:1990qb}:
\begin{eqnarray}
\lefteqn{\mathcal{L}_{ps}=
    \i \calP_{-}\left(\theta^{1}\bar{\gamma}^{-}\dot{\theta}^{1}
    +\theta^{2}\bar{\gamma}^{-}\dot{\theta}^{2}\right)
    -\i TX'^{+}\left(\theta^{1}\bar{\gamma}^{-}\dot{\theta}^{1}
    -\theta^{2}\bar{\gamma}^{-}\dot{\theta}^{2}\right)} \nonumber\\
  &&\qquad+\calP_{+}\dot{X}^{+}+\calP_{-}\dot{X}^{-}+\calP_{I}\dot{X}^{I}
    -\varphi\Big(2\calP_{+}\calP_{-}+\left(\mu^{2}X^{I}X_{I}
    +4\i\mu \theta^{1}\bar{\gamma}^{-}\Pi\theta^{2}\right)\calP_{-}^{2}\nonumber\\
  &&\qquad+\calP_{I}\calP^{I}
    +2\i T(-\calP_{-}+TX'^{-})\theta^{1}\bar{\gamma}^{-}\theta'^{1}
    +2\i T(\calP_{-}+TX'^{-})\theta^{2}\bar{\gamma}^{-}\theta'^{2}\nonumber \\
  &&\qquad+2T^{2}X'^{+}X'^{-}-T^{2}\left(\mu^{2}X^{I}X_{I}
             +4\i\mu \theta^{1}\bar{\gamma}^{-}\Pi\theta^{2}\right)
    (X'^{+})^{2}+T^{2}X'_{I}X'^{I}\Big) \nonumber \\
  &&-\rho\Big(\calP_{+}X'^{+}+\calP_{-}X'^{-}+\calP_{I}X'^{I}
    +\i(\calP_{-}-TX'^{+})\theta^{1}\bar{\gamma}^{-}\theta'^{1}+
    \i(\calP_{-}+TX'^{+})\theta^{2}\bar{\gamma}^{-}\theta'^{2}\Big). \nonumber
\end{eqnarray}
Herein, $\varphi$ and $\rho$ are Lagrange multipliers multiplying the
constraints on the canonical momenta $\calP_\mu$. Taking the limit 
$T\rightarrow 0$ in the phase space Lagrangian presents no problem
\cite{Lindstrom:1990qb}. With
the identification $V^{a}=\frac{1}{2\sqrt{\varphi}}(1,-\rho)$, we obtain
\begin{eqnarray}
  \mathcal{L}_{0}&\!=\!&V^{a}V^{b}
    \Big(2\partial_{a}X^{+}\left(\partial_{b}X^{-}
    +\i(\theta^{1}\bar{\gamma}^{-}\partial_{b}\theta^{1}
    +\theta^{2}\bar{\gamma}^{-}\partial_{b}\theta^{2})\right)
    \nonumber\\
  &&-\left(\mu^{2}X^{I}X_{I}+4\i\mu\theta^{1}\bar{\gamma}^{-}\Pi\theta^{2}\right)
    \partial_{a}X^{+}\partial_{b}X^{+}+\partial_{a}X^{I}
    \partial_{b}X_{I}\Big).
    \label{eqn:action-with-vertices}
\end{eqnarray}
The corresponding action is invariant under world-sheet diffeomorphisms with
$V^{a}$ transforming as a vector density
\begin{eqnarray}
  \delta X^{\mu} &\!=\!& \epsilon^{a}\del_{a}X^{\mu} \nonumber \\
  \delta\theta^{\alpha} &\!=\!& \epsilon^{a}\del_{a}\theta^{\alpha} \nonumber \\
  \delta V^{a} &\!=\!& -V^{b}\del_{b}\epsilon^{a}+\epsilon^{b}\del_{b}V^{a}
    +\frac{1}{2}\del_{b}\epsilon^{b}V^{a}.
\end{eqnarray}
This symmetry allows us to fix the gauge $V^{a}=(v,0)$ with $v$ a (dimensionful)
constant. The Lagrangian then becomes
\begin{eqnarray}
  \mathcal{L}_{0}&\!=v^{2}
    \Big(\!&2\dot{X}^{+}\dot{X}^{-}+\dot{X}^{I}\dot{X}_{I}
    -\left(\mu^{2}X^{I}X_{I}+4\i \mu\theta^{1}\bar{\gamma}^{-}
    \Pi\theta^{2}\right)(\dot{X}^{+})^{2}\nonumber\\
    &&+2\i\dot{X}^{+}\big(\theta^{1}\bar{\gamma}^{-}\dot{\theta}^{1}
    +\theta^{2}\bar{\gamma}^{-}\dot{\theta}^{2}\big)\Big),
    \label{eqn:action-in-diffgauge}
\end{eqnarray}
which now has to be supplemented by the constraints coming from a variation of
the action with respect to $V^{a}$. As in the tensile case there is still a
residual symmetry left after this gauge fixing. The symmetry transformations
are
\begin{eqnarray}
\delta\tau  =  f'(\sigma)\tau+g(\sigma)\mbox{,~~~~}
\delta\sigma  =  f(\sigma)
\end{eqnarray}
with $f$ and $g$ arbitrary smooth and differentiable functions of $\sigma$.
This allows us to go to light-cone gauge by choosing
$X^{+}=\frac{p^{+}}{v^{2}}\tau$. Absorbing the dependence on $v$ and $p^{+}$
into the fields according to $X^{I}\rightarrow v^{-1}X^{I}$ and
$\theta^{\alpha}\rightarrow (p^{+})^{-1/2}\theta^{\alpha}$ we obtain the
light-cone Lagrangian of the tensionless superstring in the pp-wave background
\eqref{eqn:metric}
\begin{eqnarray}
  \mathcal{L}_{0}&\!=\!&
    \dot{X}^{I}\dot{X}_{I}-m^{2}X^{I}X_{I}
    +2\i \left(\theta^{1}\bar{\gamma}^{-}\dot{\theta}^{1}
    +\theta^{2}\bar{\gamma}^{-}\dot{\theta}^{2}\right)
  -4\i m\theta^{1}\bar{\gamma}^{-}\Pi\theta^{2},
    \label{eqn:action-in-diffgauge+lcgauge}
\end{eqnarray}
where $m=\frac{\mu p^{+}}{v^{2}}$. The fields $X^{I}$ and $\theta$
are now dimensionless while $X^{+}$ and $X^{-}$ still have the dimension of
length. 
\section{Quantized solution of the model}
\label{sec:solution}
In this section we find classical solutions to and quantize the tensionless
model \eqref{eqn:action-in-diffgauge+lcgauge}. We describe the quantization
procedure and present the spectrum. 

\subsection{Solution of the classical equations of motion}
\label{subsec:solution}
The equations of motion for the action corresponding to 
\eqref{eqn:action-in-diffgauge+lcgauge} are
\begin{eqnarray}
  \ddot{X}^I + m^2 X^I=0 \label{eoma},\\
  \dot{\theta}^1- m\Pi\theta^2=0 \label{eqn:eom-theta1},\\
  \dot{\theta}^2+ m\Pi\theta^1=0 \label{eqn:eom-theta2}.
\end{eqnarray} 
Varying \eqref{eqn:action-with-vertices} with respect to $V^a$ gives the
respective ``Virasoro constraints'':
\begin{eqnarray}
  0 &\!=\!& 2p^+\dot{X}^- - m^2 X^IX_I + \dot{X}^I\dot{X}_I \nonumber\\ 
  &&\hspace{1cm}+ 2\i \big(\theta^1\bar{\gamma}^-\dot{\theta}^1 +
    \theta^2\bar{\gamma}^-\dot{\theta}^2 
    - 2m\theta^1\bar{\gamma}^-\Pi\theta^2\big) \label{eqn:LCconstraint1}\\
  0 &\!=\!& p^+X^{\prime-} + X^{\prime I}\dot{X}_I 
    + \i \big(\theta^1\bar{\gamma}^-\theta^{\prime1} +
    \theta^2\bar{\gamma}^-\theta^{\prime2} \big).\label{eqn:LCconstraint2}
\end{eqnarray}
The part containing the fermion fields in \eqref{eqn:LCconstraint1} is 
identically zero due to the equations of motion. Solving
\eqref{eoma}-\eqref{eqn:eom-theta2} for closed string boundary conditions
results in the following expansion of the transverse fields:
\begin{eqnarray}
  X^I(\sigma,\tau) &\!=\!& \cos(m\tau)x^I_0 + m^{-1}\sin(m\tau)p^I_0
    \nonumber\\
  &&+\i m^{-1}\sum_{n\neq0} \sign(n) \left\{ 
    \alpha^{1I}_n \e^{-\i(\sign(n)m\tau - 2n\sigma)} 
    +\alpha^{2I}_n \e^{-\i(\sign(n)m\tau + 2n\sigma)}\right\},
    \label{eqn:XI}\\
  \theta^1(\sigma,\tau)&\!=\!&
    \cos(m\tau)\theta^{1}_0+\sin(m\tau)\Pi\theta^{2}_0\nonumber\\
  &&+\frac{1}{\sqrt{2}}\sum_{n\neq 0}\left\{
    \theta^{1}_n\e^{-\i(\sign(n)m\tau-2n\sigma)}
    +\i\Pi\theta^{2}_n \sign(n)\e^{-\i(\sign(n)m\tau+2n\sigma)}\right\},
    \label{eqn:theta1}\\
  \theta^2(\sigma,\tau)&\!=\!&
    \cos(m\tau)\theta^{2}_0-\sin(m\tau)\Pi\theta^{1}_0\nonumber\\
  &&+\frac{1}{\sqrt{2}}\sum_{n\neq 0}\left\{
    \theta^{2}_n\e^{-\i(\sign(n)m\tau+2n\sigma)}
    -\i\Pi\theta^{1}_n \sign(n)\e^{-\i(\sign(n)m\tau-2n\sigma)}\right\}.
    \label{eqn:theta2}
\end{eqnarray}
Despite the fact that the frequencies become degenerate and, moreover, collapse
to only one frequency, these solutions look like those derived in the tensile
case \cite{Metsaev:2002re}.  Since we consider closed string boundary
conditions, we may integrate the constraint \eqref{eqn:LCconstraint2}
\begin{eqnarray}
  0 = p^+ \left(X^-(\sigma=\pi) - X^-(\sigma=0)\right)
  + \int_0^\pi\d\sigma \left\{\dot{X}^I X'_I
    +\i \left(\theta^1\bar{\gamma}^-\theta^{\prime1} +
    \theta^2\bar{\gamma}^-\theta^{\prime2} \right)\right\}, \mbox{~~~}
\end{eqnarray}
and drop the first term, because it is zero. In terms of the mode 
expansion, the constraint reads
\begin{eqnarray}
  N^1 = N^2, \hspace{2cm} N^{\cal I} \equiv 2\pi\sum_{n\neq0}
  \left\{\frac{n}{m}\,\sign(n)\,\alpha^{\cal I}_{-n}\cdot\alpha^{\cal I}_{n}
  + n \,\theta^{\cal I}_{-n}\bar{\gamma}^-\theta^{\cal I}_n  \right\},
\end{eqnarray}
with ${\cal I}=1,2$. This is the usual level-matching condition.

\subsection{Light-cone Hamiltonian}
\label{sec:LCH}
The light-cone Hamiltonian is derived by integrating the momentum conjugate 
to the time coordinate $X^+$, \ie
\begin{eqnarray}
  {H}_{lc} = -\int\d\sigma \,\calP^-.
\end{eqnarray}
In the diffeomorphism and light-cone gauge and by using 
\eqref{eqn:eom-theta1}-\eqref{eqn:LCconstraint1} it
can be written as:
\begin{eqnarray}
  {H}_{lc}= \frac{v^2}{p^+}\int\d\sigma \left\{ \dot{X}^I\dot{X}_I 
  + m^2 X^IX_I
  + 2\i\big(\theta^1\bar{\gamma}^-\dot{\theta}^1 
  + \theta^2\bar{\gamma}^-\dot{\theta}^2\big)\right\}.
\end{eqnarray}
In terms of the mode expansions of the fields we find 
\begin{eqnarray}
  {H}_{lc}&\!=\!&\frac{\pi v^2}{p^+}\left(p_0^2 +m^2x_0^2\right)
    + 4\pi\i\frac{m v^2}{p^+}\theta_0^1\bar{\gamma}^-\Pi\theta_0^2 \nonumber\\ 
  &&~+\frac{2\pi v^2}{p^+}\sum_{n\neq0}\!
    \Big\{\alpha_{-n}^{1}\cdot\alpha_{n}^{1} 
    + \alpha_{-n}^{2}\cdot\alpha_{n}^{2} 
    +m\,\sign(n)\left(\theta^1_{-n}\bar{\gamma}^-\theta^1_n 
    + \theta^{\:2}_{-n}\bar{\gamma}^-\theta^{\:2}_n\right)\!\!\Big\}.
    \mbox{~~~~~}
  \label{eqn:LCHamiltonian}
\end{eqnarray}
This light-cone Hamiltonian gives the energy for the 
classical states.

\subsection{Canonical quantization}
\label{sec:quantization}
To obtain the corresponding quantum theory, we apply the equal time canonical
commutation relations to the fields $X^I$ and their conjugate momenta
$\calP^I=2\dot X^I$:
\begin{eqnarray}
  \left[\calP^I(\sigma,\tau),X^J(\sigma',\tau)\right] = 
  -\i\delta^{IJ}\delta(\sigma-\sigma').
\end{eqnarray}
These translate into commutation relations for the modes as follows
\begin{eqnarray}
  \left[p^I_0,x^J_0\right] &\!=\!& -\frac{\i}{2\pi}\delta^{IJ},
   \label{eqn:commpx}\\ \left[\alpha^{{\calI}I}_{n},
   \alpha^{{\cal J} J}_{n'}\right]&=&
    \frac{m}{4\pi} \,\sign(n)\,\delta_{n+n',0}\,
    \delta^{IJ}\delta^{{\calI}{\calJ}},~n,n'\in\Z\backslash\{0\}\label{eqn:commaa}.
\end{eqnarray}
For the fermionic fields we have a second class constraint, since the momenta
for these fields, $-2\i\bar{\gamma}^-\theta^{\cal I}$, are linearly dependent
on the fields themselves. This yields the anti-commutation relations
\begin{eqnarray}
  \big\{\theta^{{\calI}\alpha}(\sigma,\tau),
         \theta^{{\calJ}\beta}(\sigma',\tau)\big\}&=&
    \frac{1}{8\pi}(\gamma^+)^{\alpha\beta}\delta^{\calI\calJ} 
    \delta(\sigma - \sigma'),
\end{eqnarray}
which for the modes results in 
\begin{eqnarray}
  \big\{\theta^{{\calI}\alpha}_n,\theta^{{\calJ}\beta}_{n'}\big\} &\!=\!&
    \frac{1}{8\pi}(\gamma^+)^{\alpha\beta}\delta^{\calI\calJ}\delta_{n+n',0},
    ~n,n'\in\Z.\label{eqn:commtt}
\end{eqnarray}
Equations \eqref{eqn:commaa}, \eqref{eqn:commtt} imply that we can promote the
bosonic and fermionic negative frequency modes to creation operators and the
corresponding positive ones to annihilation operators. To define a proper
vacuum state, we collect the zero-modes into harmonic oscillator modes
\begin{eqnarray}
  \alpha_0^I={\textstyle \frac{m}{2}}
    \big(x_0^I+{\textstyle \frac{\i}{m}}p_0^I\big),
  \hspace{0.5cm}
  \alpha_0^{\dagger I}={\textstyle \frac{m}{2}}
    \big(x_0^I-{\textstyle \frac{\i}{m}}p_0^I\big),
\end{eqnarray}
obeying the commutation relation
\begin{eqnarray}
  &\big[\alpha_0^{I},\alpha_0^{\dagger J}\big] = \frac{m}{4\pi}\delta^{IJ}.
\end{eqnarray}
For the vacuum state we require that 
\begin{eqnarray}
  \alpha^{\calI}_n \left|0;p^+\right\rangle = 0,~n\geq 0, \mbox{~~~~}
  \theta^{\calI}_n \left|0;p^+\right\rangle = 0,~n>0. 
  \label{eqn:vacuum}
\end{eqnarray}
While there is a unique choice for the bosonic part of the vacuum state, there
is a variety of four possibilities for choosing the fermionic part
\cite{Metsaev:2002re}, but in the end all of the latter are equivalent by a
redefinition of the $\theta$-modes. This will be of no importance for what
follows, our analysis can indeed be carried out for any of the possible choices
of the fermionic vacuum states.  The light-cone Hamiltonian of the quantized
theory can be obtained by using the above commutation relations to
normal-order the result \eqref{eqn:LCHamiltonian}: 
\begin{eqnarray}
  {H}_{lc}&\!=\!&
    \frac{4\pi v^2}{p^+} \alpha_0^{\dagger I}\alpha_0^J \delta_{IJ} 
    + 4\pi\i \frac{m v^2}{p^+}\theta_0^1\bar{\gamma}^-\Pi\theta_0^2 \nonumber\\ 
  &&+\frac{4\pi v^2}{p^+}\sum_{n>0}\!\Big\{\alpha_{-n}^{1}\cdot\alpha_{n}^{1}
    + \alpha_{-n}^{2}\cdot\alpha_{n}^{2} 
    +m\,\left(\theta^1_{-n}\bar{\gamma}^-\theta^1_n 
    + \theta^{\:2}_{-n}\bar{\gamma}^-\theta^{\:2}_n\right)\!\!\Big\},
  \label{eqn:Hlc}
\end{eqnarray}
where we used that in $D=10$ the spinor space is 16 dimensional and the
$\kappa$-symmetry fixing condition to drop the term proportional to
$\gamma^-\bar{\gamma}^+$. This result resembles the light-cone Hamiltonian in
\cite{Metsaev:2002re} despite the frequency degeneracy already seen for the
fields. This implies that the energy spectrum of the quantum theory is
infinitely degenerate. 

\subsection{The spectrum and the space of states}
\label{sec:spectrum}
Starting from the vacuum state \eqref{eqn:vacuum}, we count the number of
states obtained by exciting with the bosonic oscillator modes on every energy
level. The states are written in terms of the oscillators that act on
the vacuum. For the first levels, we also give the corresponding
representations under the $SO(4)\times SO^\prime(4)$ symmetry.
\begin{eqnarray}
  \begin{array}{|@{\quad}l@{\quad}|c|@{\quad}l@{\quad}|}
    \hline
    \mbox{state}&\mbox{energy-}&\mbox{number of states}\\
    &\mbox{level}& \\ \hline
    \bb{1}&0&(1,1)\\ \hline
    \alpha_0^{\dagger i},~\alpha_0^{\dagger i'}&1&(4,1)\oplus(1,4)\\ \hline
    
    \alpha_0^{\dagger i}\alpha_0^{\dagger j}&2&(1,1)\oplus(9,1)\\
    \alpha_0^{\dagger i'}\alpha_0^{\dagger j'}&&(1,1)\oplus(1,9)\\
    \alpha_0^{\dagger i}\alpha_0^{\dagger i'}&&(4,4)\\
    \alpha^{1i}_{-n}\alpha^{2j}_{-n}&&
      (1,1)\oplus(9,1)\oplus(3^+,1)\oplus(3^-,1),~n\in\N\\
    \alpha^{1i'}_{-n}\alpha^{2j'}_{-n}&&
      (1,1)\oplus(1,9)\oplus(1,3^+)\oplus(1,3^-),~n\in\N\\
    \alpha^{1i}_{-n}\alpha^{2i'}_{-n}&&(4,4),~n\in\N\\
    \hline
    (\alpha^{\dagger}_0)^3&3&120=
      (1,1)\oplus(19,1)\oplus(4,1)\oplus(4,9)\:\oplus\\
    &&\phantom{120=\:}(9,4)\oplus (1,4)\oplus(1,19)\oplus(1,1)\\
    \alpha^\dagger_0\alpha^1_{-n}\alpha^2_{-n}&&512,~n\in\N\\
    \alpha^1_{-n}\alpha^1_{-m}\alpha^2_{-(n+m)},\,\alpha^1\alpha^2\alpha^2&&
      288+288,~n,m\in\N\\ 
    \hline
    \hspace*{2cm}\vdots&\vdots&\hspace*{3cm}\vdots \\ \hline
  \end{array}\nonumber
\end{eqnarray}
The spectrum built from the fermionic oscillators reads:
\begin{eqnarray}
  \begin{array}{|@{\quad}l@{\quad}|@{\quad}c@{\quad}|@{\quad}l@{\quad}|}
    \hline
    \mbox{state}&\mbox{energy}&\mbox{number of states}\\ \hline
    \bb{1}&0&1\\ \hline
    \theta^{\calI\alpha}_0&1&2\cdot 16=32\\ \hline
    \theta^{\calI\alpha}_0\theta^{\calJ\beta}_0&2&496=256+240\\
    \epsilon_{\calI\calJ}\theta^{\calI\alpha}_{-n}\theta^{\calJ\beta}_{-n}&& 
      256,~n\in \N\\ 
    \hline
    \theta^{\calI\alpha}_0\epsilon_{\calJ\calK}\theta^{\calJ\beta}_{-n}
    \theta^{\calK\gamma}_{-n}&3&n\in \N\\
    \epsilon_{\calI\calJ}
    \theta^{\calI\alpha\vphantom{\gamma}}_{-n\vphantom{()}}
    \theta^{\calI\beta\vphantom{\gamma}}_{-m\vphantom{()}}
    \theta^{\calJ\gamma}_{-(n+m)}&&n,m\in \N\\ \hline
    \hspace*{1.5cm}\vdots&\vdots&\hspace*{1.5cm}\vdots \\ \hline
  \end{array}\nonumber
\end{eqnarray}
It is further possible to build states out of both the bosonic and the 
fermionic oscillators, but we do not pursue this here.
\subsection{Degeneracy of the bosonic spectrum at the second energy level}
From the above tables one sees that the bosonic part of the spectrum at
energy level 2 is generated by states of the form 
$T_{IJ}\alpha_{-n}^{1I}\alpha_{-n}^{2 J}\ket{0}$, where 
$\alpha_{-0}^{\calI I}\equiv \alpha_0^{\dagger I}$, 
$\alpha_{0}^{\calI I}\equiv \alpha_0^I$ and $T_{IJ}$ is some polarization
tensor. This corresponds to exactly one excitation in the $n$-th pair of
oscillators. Denoting these states by the quantum number $n$, \ie
\begin{eqnarray}
  \ket{n}_{IJ}=\frac{2\pi}{m}\alpha_{-n}^{1 I}\alpha_{-n}^{2 J}\ket{0;p^+},
\end{eqnarray}
we define two operators\footnote{The additional factor in 
      ${\cal A}^+$ is necessary because there is only one
      set of zero-modes, while all higher oscillator modes come in pairs.}
\begin{eqnarray}
  {\cal A}^+&\!=\!&\left(\frac{2\pi}{m}\right)^2\sum_{IJ}\sum_{n\ge 0}
    \sqrt{n+1}\alpha_{-n-1}^{1I}\alpha_{n}^{1I}
    \alpha_{-n-1}^{2J}\alpha_{n}^{2J}
      \cdot\big(1-{\textstyle \frac{1}{2}}\delta_{n,0}\big) \\
  {\cal A}^-&\!=\!&\left(\frac{2\pi}{m}\right)^2\sum_{IJ}\sum_{n\ge 0}
    \sqrt{n}\alpha_{-n+1}^{1I}\alpha_{n}^{1I}\alpha_{-n+1}^{2J}\alpha_{n}^{2J}.
\end{eqnarray}
These operators commute to $\id$ on the states under consideration and act as
usual harmonic oscillator raising and lowering operators:
\begin{eqnarray}
  {\cal A}^+\ket{n}_{IJ}=\sqrt{n+1}\ket{n}_{IJ},~~
  {\cal A}^-\ket{n}_{IJ}=\sqrt{n}\ket{n-1}_{IJ},~~
  {\cal A}^-\ket{0}_{IJ}=0.
\end{eqnarray}
Since there exists only one set of zero-modes, $\ket{0}_{IJ}$ is automatically
symmetrized. The harmonic oscillator Hamiltonian is defined in the usual way as
well, \ie $\tilde{H}={\cal A}^+{\cal A}^-$, yielding 
$\tilde{H}\ket{n}=n\ket{n}$. This algebra obviously commutes with the
light-cone Hamiltonian ${H}_{lc}$ \eqref{eqn:Hlc}. Thus its action does
not change the energy but rather reflects the degeneracy of the states.

\subsection{Degeneracy of the bosonic spectrum at energy level 3}
The discussion for energy level 3 is carried out in the same way was done for
level 2. Denoting the states by 
\begin{eqnarray}
  \ket{n,m}_{IJK}^\calI\equiv\alpha_{-n}^{\calJ I} \alpha_{-m}^{\calJ J}
  \alpha_{-(n+m)}^{\calI K}\ket{0;p^+},
    ~\calI\neq\calJ,
\end{eqnarray}
we may write down the following series (for $n+m=\mbox{even}$)
\begin{eqnarray}
  \ket{n,n}^1
  \rightarrow \ket{n-1,n+1}^1
  \rightarrow \ldots
  \rightarrow \ket{1,2n-1}^1 \hspace*{3cm}\nonumber \\
  \rightarrow \ket{0,2n}^{1=2}
  \rightarrow \ket{1,2n-1}^2
  \rightarrow \ldots
  \rightarrow \ket{n,n}^2,
\end{eqnarray}
and correspondingly for odd $n+m$
\begin{eqnarray}
  \ket{n-1,n}^1
  \rightarrow \ket{n-2,n+1}^1
  \rightarrow \ldots
  \rightarrow \ket{1,2n-2}^1 \hspace*{3cm}\nonumber \\
  \rightarrow \ket{0,2n-1}^{1=2}
  \rightarrow \ket{1,2n-2}^2
  \rightarrow \ldots
  \rightarrow \ket{n-1,n}^2.
\end{eqnarray}
It is easy to see that these states form integer spin representations of
$su(2)$ of increasing level as $n+m$ gets bigger. But these are nothing but the
different levels of an affine $\widehat{su}(2)$ algebra! Correspondingly, the
Heisenberg algebra in the previous subsection is indeed an affine $\hat{u}(1)$.

\subsection{Bosonic sector partition function}
Generalizing the above results, we obtain a partition function $Z_E$ for each
energy level $E$ by the recursive formula
\begin{eqnarray}
  Z_0(x,y)&=&1,\\
  Z_E(x,y)&=&Z_{E-1}+\sum_{k=1}^{E-1}q_k(x)q_{E-k}(y)=
    \sum_{mn}\alpha_{mn}x^m y^n,
\end{eqnarray}
where $q_k(x)=\sum_{n=0}^{\infty}q_k^n x^n$ is the generating functional for
the number of partitions of $n$ with exactly $k$ parts. The sub-series with
coefficients $\alpha_{nn}$ then encodes the dimension of the $n$-th multiplet at
energy level $E$. As $E\rightarrow\infty$, the partition function goes to a
very simple limit, namely
\begin{eqnarray}
  Z_E(x,y)\rightarrow Z_\infty(x,y)=
  \frac{(xy)^{1/24}}{\eta(x)\eta(y)}-\frac{x^{1/24}}{\eta(x)}
  -\frac{y^{1/24}}{\eta(y)},
\end{eqnarray}
where $\eta(x)$ is the Dedekind $\eta$-function. In this context, an 
interesting question is the effect of this huge degeneracy on the BMN 
correspondence \cite{Berenstein:2002jq}. In particular, vanishing tension is
related to the weak-coupling limit of the corresponding conformal field
theory. Another open question is the relation to higher spin field theories. 

\section{The tensionless string as a limit of ordinary string theory
in a pp-wave background}
\label{sec:limit}
In this section we reverse the order of the procedure discussed in section
\ref{sec:solution}. We first recapitulate some details of the quantized tensile
theory in the pp-wave background \eqref{eqn:metric} \cite{Metsaev:2002re} and
then take the limit $T\rightarrow 0$ within this setting in a certain way. We
show that this limit recovers our previous results. The light-cone and
$\kappa$-symmetry fixed Lagrangian in the tensile case reads
\begin{eqnarray}
  \tilde{\cal L}&\!=\!&\tilde{\cal L}_B+\tilde{\cal L}_F \nonumber \\
  &\!=\!&(\del_+\tilde X^I\del_-\tilde X_I-\tilde m^2\tilde X_I^2)
     +2\i (\tilde\theta^1\bar\gamma^-\del_+\tilde\theta^1
     +\tilde\theta^2\bar\gamma^-\del_-\tilde\theta^2
     -2\tilde m\tilde\theta^1\bar\gamma^-\Pi\tilde\theta^2).
\end{eqnarray}
Here, tilde refers to quantities in the tensile theory to distinguish them from
the ``bare'' tensionless quantities.  As in the previous discussion, we absorb
the string tension $T$ in a rescaling of the fields.  The equations of motions
yield the closed string solutions
\begin{eqnarray}
  \tilde X^I(\sigma,\tau)&\!=\!&
    \cos(\tilde m\tau)\tilde x^I_0+\frac{1}{\tilde m}
    \sin(\tilde m\tau)\tilde p^I_0\nonumber \\ 
  &&+\i\sum_{n\neq 0}\frac{1}{\tilde \omega_n}\big\{
    \tilde\alpha_n^{1I}\e^{-\i(\tilde\omega_n\tau-2n\sigma)}
    +\tilde\alpha_n^{2I}\e^{-\i(\tilde\omega_n\tau+2n\sigma)}\big\},\\
  \tilde\theta^{1}(\sigma,\tau)&\!=\!& \cos(\tilde m\tau)\tilde\theta^{1}_0+
    \sin(\tilde m\tau)\Pi\tilde\theta^{2}_0\nonumber \\ 
  &&+\sum_{n\neq 0} \tilde c_n\big\{
    \tilde\theta^{1}_n\e^{-\i(\tilde \omega_n\tau-2n\sigma)}
    +\i\Pi\tilde\theta^{2}_n\frac{\tilde \omega_n-2 n}{\tilde m}
    \e^{-\i(\tilde \omega_n\tau+2n\sigma)}\big\},
\end{eqnarray}
and similar for $\tilde\theta^2$. Herein, 
\begin{eqnarray}
 \tilde\omega_n=\mbox{sign}(n)\sqrt{\tilde m^2+4n^2}\mbox{~~~and~~~}
 \tilde c_n=
    \frac{1}{\sqrt{1+\left(\frac{\tilde\omega_n-2 n}{\tilde m}\right)^2}}.
\end{eqnarray}
In taking the tension to zero, we set $T=v^2\lambda$ and take the
dimensionless quantity $\lambda\rightarrow 0$. This limit is unproblematic if
accompanied by a rescaling of the world-sheet time 
$\tau\rightarrow \bar\tau=\frac{\tau}{\lambda}$ where $\bar\tau$ is kept fixed.
This corresponds to keeping the light-cone coordinate $\tilde{X}^+$ fixed in
this limit and results in the world-sheet becoming a null surface. We introduce
the following set of redefinitions:
\begin{eqnarray}
  m&\!=\!&\lambda\tilde m \nonumber\\
  \omega_n&\!=\!&\lambda\tilde\omega_n 
    =\sign(n)\sqrt{m^2+4\lambda^2 n^2}\nonumber\\
  x^I_0&\!=\!&\frac{1}{\sqrt{\lambda}}\tilde x^I_0\nonumber\\
  p^I_0&\!=\!&\sqrt{\lambda}\tilde p^I_0\nonumber\\
  \alpha^I_n&\!=\!&\sqrt{\lambda}\tilde\alpha^I_n. \label{eqn:rescalings}
\end{eqnarray}
The new oscillator modes are independent of the string tension. This
implies
\begin{eqnarray}
  \tilde X^I(\sigma,\bar\tau)&\!=\!&\sqrt{\lambda}\Big[
    \cos(m\bar\tau)x^I_0+\frac{1}{m}
    \sin(m\bar\tau)p^I_0\nonumber \\ 
  &&+\i\sum_{n\neq 0}\frac{1}{\omega_n}\big\{
    \alpha_n^{1I}\e^{-\i(\omega_n\bar\tau-2n\sigma)}
    +\alpha_n^{2I}\e^{-\i(\omega_n\bar\tau+2n\sigma)}\big\}\Big].
\end{eqnarray}
Because we absorbed the string tension into the fields, we may now take
$\lambda\rightarrow 0$ in a non-trivial way after introducing new coordinates 
$X^I(\sigma,\bar\tau)={\lambda}^{-1/2}\tilde X^I(\sigma,\bar\tau)$. 
Comparing to \eqref{eqn:XI} we find a perfect correspondence. Already at this
stage (just looking at the bosonic part of the theory) this indicates that - in
contrast to flat space - tensionless string theory in the pp-wave is an
uncomplicated limit of the ordinary tensile one. To underpin this statement we
take a closer look at the fermionic fields $\tilde\theta^1(\sigma,\tau)$ and
$\tilde\theta^2(\sigma,\tau)$. The fermionic oscillator modes are independent
of the string tension even in the tensile theory and thus, under
$\lambda\rightarrow 0$, we may take 
$\tilde\theta^{{\cal I}\alpha}_n\rightarrow\theta^{{\cal I}\alpha}_n$.
Furthermore, $c_n\rightarrow\frac{1}{\sqrt{2}}$, 
$\frac{\omega_n-2\lambda n}{m}\rightarrow \sign(n)$ for all $n$. In this limit
the fields become
\begin{eqnarray}
  \theta^1(\sigma,\bar\tau)&\!=\!&
    \cos(m\bar\tau)\theta^{1}_0+\sin(m\bar\tau)\Pi\theta^{2}_0\nonumber\\
  &&+\frac{1}{\sqrt{2}}\sum_{n\neq 0}\left\{
    \theta^{1}_n\e^{-\i(\sign(n)m\bar\tau-2n\sigma)}
    +\i\Pi\theta^{2}_n \sign(n)\e^{-\i(\sign(n)m\bar\tau+2n\sigma)}\right\},
    \label{eqn:thetalimit}\\
  \theta^2(\sigma,\bar\tau)&\!=\!&
    \cos(m\bar\tau)\theta^{2}_0-\sin(m\bar\tau)\Pi\theta^{1}_0\nonumber\\
  &&+\frac{1}{\sqrt{2}}\sum_{n\neq 0}\left\{
    \theta^{2}_n\e^{-\i(\sign(n)m\bar\tau+2n\sigma)}
    -\i\Pi\theta^{1}_n \sign(n)\e^{-\i(\sign(n)m\bar\tau-2n\sigma)}\right\},
\end{eqnarray}
again a perfect match with \eqref{eqn:theta1}, \eqref{eqn:theta2}. We can even go
one step further and deduce the tensionless supersymmetric Lagrangian from the
tensile one:\footnote{For the bosonic part of the Lagrangian, similar
considerations have been carried out in the context of string field theory 
\cite{Chu:2002mg}.} The bosonic part of the light-cone Lagrangian
\cite{Metsaev:2002re} reads
\begin{eqnarray}
  \tilde{\cal L}_B=\del_\tau\tilde X^I\del_\tau\tilde X^I-
    \del_\sigma\tilde X^I\del_\sigma\tilde X^I-
    \tilde m^2\tilde X_I^2.
\end{eqnarray}
Applying the rescalings \eqref{eqn:rescalings}, we find
\begin{eqnarray}
  \tilde{\cal L}_B=\frac{1}{\lambda}\left(\del_{\bar\tau} X^I
  \del_{\bar\tau}X^I-\lambda^2\del_\sigma X^I\del_\sigma X^I-m^2 X_I^2\right).
\end{eqnarray}
After identifying ${\cal L}_B=\lambda\tilde{\cal L}_B$ (coming from the
combined scaling of $\tau$ which leaves the action invariant), we can take 
the limit $\lambda\rightarrow 0$ on ${\cal L}_B$ to obtain
\begin{eqnarray}
  {\cal L}_B=\dot X_I^2-m^2 X_I^2,
  \label{eqn:LCGLagrangian}
\end{eqnarray}
If we apply this same procedure to the fermionic part of the Lagrangian, we
derive
\begin{eqnarray}
  \tilde{\cal L}_F&\!=\!&2\i\Big(\tilde\theta^1\bar\gamma^-
    (\del_\tau+\del_\sigma)\tilde\theta^1
    +\tilde\theta^2\bar\gamma^-(\del_\tau-\del_\sigma)\tilde\theta^2
    -2\tilde m\tilde\theta^1\bar\gamma^-
    \Pi\tilde\theta^2\Big) \nonumber \\
  &\!=\!&\frac{2\i}{\lambda}\Big(
    \theta^1\bar\gamma^-(\del_{\bar\tau}+\lambda\del_\sigma)\theta^1
    +\theta^2\bar\gamma^-(\del_{\bar\tau}-\lambda\del_\sigma)\theta^2
    -2m\theta^1\bar\gamma^-\Pi\theta^2\Big).\label{eqn:L-fermionic}
\end{eqnarray}
With ${\cal L}_F=\lambda\tilde{\cal L}_F$, the limit results in:
\begin{eqnarray}
  {\cal L}_F=2\i\left(\theta^1\bar\gamma^-\del_{\bar\tau}\theta^1
  +\theta^2\bar\gamma^-\del_{\bar\tau}\theta^2
  -2m\theta^1\bar\gamma^-\Pi\theta^2\right).
\end{eqnarray}
But then, ${\cal L}={\cal L}_B+{\cal L}_F$ is the light-cone Lagrangian we
arrived at in \eqref{eqn:action-in-diffgauge+lcgauge}. 

We now take a closer look at the behavior of the commutation relations of the
tensile quantum theory in this limit:
\begin{eqnarray}
  \left[\tilde{p}_0^I,\tilde{x}_0^J\right]=-\frac{\i}{2\pi}\delta^{IJ}
  &\longrightarrow&\left[p_0^I,x_0^J\right]=-\frac{\i}{2\pi}\delta^{IJ}
   \nonumber\\
  \left[\tilde{\alpha}_{n}^{{\calI}I},\tilde{\alpha}_{n'}^{{\calJ}J}\right]=
   \frac{1}{4\pi}\tilde{\omega}_{n}\delta_{n+n',0}\delta^{IJ}
   \delta^{{\calI}{\calJ}}
  &\longrightarrow& 
  \left[\alpha_{n}^{{\calI}I},\alpha_{n'}^{{\calJ}J}\right]=
    \frac{m}{4\pi}\sign(n)\delta_{n+n',0}\delta^{IJ}
    \delta^{{\calI}{\calJ}}\mbox{~~~~~~~}\\
  \big\{\tilde{\theta}_{n}^{{\calI}\alpha},
    \tilde{\theta}_{n'}^{{\calJ}\beta}\big\}=
    \frac{1}{8\pi}\left(\gamma^+\right)^{\alpha\beta}
    \delta_{n+n',0}\delta^{\calI\calJ}
  &\longrightarrow&
  \big\{\theta_{n}^{{\calI}\alpha},\theta_{n'}^{{\calJ}\beta}\big\}=
    \frac{1}{8\pi}\left(\gamma^+\right)^{\alpha\beta}
    \delta_{n+n',0}\delta^{\calI\calJ}.
    \nonumber
\end{eqnarray}
This is in perfect agreement with (\ref{eqn:commpx}, \ref{eqn:commaa},
\ref{eqn:commtt}). By glancing at the spectrum of the quantized theory, we
learn that this continues in a straight-forward way, which shows that
tensionless string theory in the pp-wave background \eqref{eqn:metric} is a
well-defined limit of ordinary string theory in this context. In flat space,
as a contrast, tensionless string theory can be defined by quantizing the
classical tensionless Lagrangian, but it seems very difficult to obtain it from
the quantized theory. 

\section{Homogeneous plane waves}
\label{sec:hppwaves}
It is an interesting question whether the presented considerations extend
to a more general class of plane wave backgrounds or if they are an intrinsic
property of the pp-wave background \eqref{eqn:metric}. For this purpose, this
section treats the case of a homogeneous plane wave. The most general smooth
such background can be described by a metric containing two constant matrices
$k_{IJ}$ and $f_{IJ}$ \cite{Blau:2002js}:
\begin{eqnarray}
  \d s^2=2\dx^+\dx^-+k_{IJ}x^Ix^J{\dx^+}\dx^++2f_{IJ}x^I\dx^J\dx^++\dx^I\dx_I.
  \label{eqn:Hppwavemetric}
\end{eqnarray}
Obviously, in general, such a background is no longer maximally supersymmetric.
In this section, we allow for an arbitrary dimension $D$, such that indices
$I,J,K$ denote $D-2$ transverse coordinates. Additionally, we allow for a
$B$-field given by the $D-2$ non-vanishing components $B_{I+}=h_{IJ}x^J$.  Not
surprisingly, in $D=10$ dimensions, and for $h,f=0$ and
$k_{IJ}=-\mu^2\delta_{IJ}$, we get back \eqref{eqn:metric}. By a rotation of
the transverse coordinates, $k$ can be chosen to be diagonal:
$k_{IJ}=k_I\delta_{IJ}$. The resulting string theory is an integrable model and
was solved by Blau et.\,al.\ \cite{Blau:2003rt} via a frequency base ansatz:
\begin{eqnarray}
  X^I(\sigma,\tau)=\sum_{n=-\infty}^{\infty}X_n^I(\tau)\e^{2\i n\sigma},~~~
  X_n^I(\tau)=
    \sum_{\ell=1}^{2D}\xi_{n\ell\,}^{\vphantom{I}}
    a_{n\ell\,}^{I}\e^{\i \omega_{n\ell\,}\tau}.
\end{eqnarray}

In the quantized theory, $\xi_{n\ell\,}$ will become the oscillator modes. The
$a_{n\ell\,}^{I}$ are vectors given by the eigen directions of the following
matrix:
\begin{eqnarray}
  M_{IJ}(\omega,n)=
    (\omega^2+k_I-4T^2n^2)\delta_{IJ}+2\i\omega f_{IJ}+4\i Tnh_{IJ}.
  \label{eqn:matrix}
\end{eqnarray}
The allowed frequencies are then determined by $\det M(\omega_{n\ell\,},n)=0$
while the eigen directions are given by
$M_{IJ}(\omega_{n\ell\,}^{\vphantom{I}},n)a_{n\ell\,}^{I}=0$.  It is easy to
see that for $h,f=0$ and constant $k_I=-\mu^2$, one obtains
$\omega_n^2-\mu^2-4T^2n^2=0$ or - equivalently -
$\omega_{n\pm}=\pm\sqrt{\mu^2+4T^2n^2}$ as expected. We previously obtained the
frequencies in the tensionless case by letting $T\rightarrow 0$ directly in
the corresponding expression. Looking at \eqref{eqn:matrix}, we find that we
may do the same here and we again find that the frequencies become degenerate
and equal to the frequencies for $n=0$:
\begin{eqnarray}
  \omega_{n\ell\,}\rightarrow \omega_\ell,
  ~~~ \det M(\omega_\ell)\equiv \det M(\omega_\ell,0) = 0.
\end{eqnarray}
From this short remark, we learn that the huge energy degeneration in the space
of states is not a special property of the previously discussed maximally
supersymmetric pp-wave background but occurs also for more general plane waves.
It seems plausible that this result still holds after going over to the 
quantum theory here as well.

\section{Conclusion}\label{sec:conclusion}
We have presented the quantized theory of a tensionless superstring in the
the constant plane wave background. This theory is a limit of the quantized
tensile superstring. In particular, this implies that the tensionless
limit can be consistently obtained in the quantized theory of tensile strings
in such pp-waves. We find this an interesting property of these backgrounds that 
seems to be related to the existence of a dimensionful parameter. As discussed
in a series of articles in the past, this seems not to be the case in flat
space.  

Since the string tension and background mass term always come in the combination
$\frac{m}{T}$, we conclude that a tensionless limit for fixed $m$ is
equivalent to keeping the tension fixed and letting $m\rightarrow\infty$, 
corresponding to an infinitely curved space.

Interesting open questions prompted by our investigation are those related
to the BMN limit discussed above, a generalization to similar backgrounds
such as the homogeneous or even more general plane waves, see, \eg 
\cite{Bredthauer:2003fs}, and (again) the relation to higher spin theories.

\acknowledgments
The authors are grateful to I.~Bakas, B.~Sundborg, P.~Sundell and A.~Zheltukhin
for illuminating discussions and helpful suggestions. We would also like to
thank the referee for useful comments. The research of UL is supported in part
by VR grant 650-1998368.

\appendix

\section{Notation} \label{app:notation}
Conventions for the indices:\\[1ex]
\begin{tabular}{@{\qquad}ll}
$\mu,\nu = 0,1,\ldots,9$ & $SO(9,1)$ vector indices\\
$I,J     = 1,\ldots,8$  & $SO(8)$ vector indices\\
$i,j     = 1,\ldots,4$  & $SO(4)$ vector indices\\
$i',j'   = 5,\ldots,8$  & $SO^\prime(4)$ vector indices\\
$\alpha,\beta,\gamma = 1,\ldots,16$ & $SO(9,1)$ spinor indices\\
$\calI, \calJ = 1,2$ & labels for the two sets of oscillators\\
$a,b     = 0,1 = \tau,\sigma$ & 2D world-sheet coordinate indices\\
\end{tabular}
\\

The chiral representation for the $32\times32$ Dirac matrices $\Gamma^\mu$ are
given by the off-diagonal $16\times16$ blocks $\gamma^\mu$'s of the
$\Gamma^\mu$'s,
\begin{eqnarray}
  \Gamma^\mu = \left(
    \begin{array}{cc}
      0&\gamma^\mu\\
      \bar{\gamma}^\mu&0
    \end{array}\right).
\end{eqnarray}
The $\gamma^\mu$ satisfy $\bar{\gamma}^\mu = \gamma^\mu_{\alpha\beta}$ and 
\begin{eqnarray}
  \gamma^\mu \bar{\gamma}^\nu + \gamma^\nu\bar{\gamma}^\mu = 2\eta^{\mu\nu}.
\end{eqnarray} 
The $\gamma^\mu$ matrices are real and symmetric. The fermionic mass operators
$\Pi$ and $\bar\Pi$ satisfying $\Pi^2=1$ are given by
\begin{eqnarray}
  \Pi^\alpha_{\phantom{\alpha}\beta} \equiv 
    (\gamma^1\bar{\gamma}^2\gamma^3
     \bar{\gamma}^4)^\alpha_{\phantom{\alpha}\beta},
  \hspace{1cm} 
  \Pi^{\prime\alpha}_{\phantom{\alpha}\beta} \equiv 
    (\gamma^5\bar{\gamma}^6\gamma^7
     \bar{\gamma}^8)^\alpha_{\phantom{\alpha}\beta}\\
  \bar{\Pi}^\alpha_{\phantom{\alpha}\beta} \equiv 
    (\bar{\gamma^1}\gamma^2
     \bar{\gamma^3}\gamma^4)^\alpha_{\phantom{\alpha}\beta},
  \hspace{1cm}
  \bar{\Pi}^{\prime\alpha}_{\phantom{\alpha}\beta} \equiv 
    (\bar{\gamma}^5\gamma^6
     \bar{\gamma}^7\gamma^8)^\alpha_{\phantom{\alpha}\beta}
\end{eqnarray}
The identity $(\gamma^0\bar{\gamma}^9)^2=1$ implies
\begin{eqnarray}
  \gamma^0\bar{\gamma}^9\gamma^\pm = \pm\gamma^\pm,
  \mbox{~~~}
  \bar\gamma^\pm\gamma^0\bar{\gamma}^9 = \mp\bar\gamma^\pm,
\end{eqnarray}
allowing us to define the $\gamma^\pm$, $\bar\gamma^\pm$ as eigenmatrices of
$\gamma^0\bar{\gamma}^9$. For a more detailed prescription we refer to
\cite{Metsaev:2001bj,Metsaev:2002re}.

\section{The SUSY-algebra}
\label{sec:SUSY}
The symmetries of the supersymmetric action can be found following the lines in
\cite{Metsaev:2001bj}: The $\kappa$-symmetry fixed Lagrangian
\eqref{eqn:action-in-diffgauge} in complex $\theta,\bar\theta$
notation, where $\theta=\theta^1+\i\theta^2$ reads
\begin{eqnarray}
  {\cal L}=\left(2\dot X^+\dot X^--X_I^2(\dot X^+)^2
    +\dot X_I^2\right)+2\i\dot X^+\left(\bar\theta\bar\gamma^-
    \dot\theta+\theta\bar\gamma^-\dot{\bar\theta}
    +2\i\dot X^+\bar\theta\bar\gamma^-\Pi\theta\right).~~~~~
\end{eqnarray}
Here, $v$ and $\mu$ are gauge fixed to $1$, but can be introduced as in the
previous sections.  Since we did not change the background in comparison to
Metsaev, the (infinitesimal) transformations giving the symmetries of the
Lagrangian should be the same. Indeed, translations in the $X^-$-plane 
($\delta X^-=\epsilon^-$) give rise to a conserved current 
${{\cal P}^{+a}=2\dot X^+\delta^{a 0}}$, where $a=(0,1)$ are the directions on
the world-sheet. It is then easy to check that the conserved kinematical
currents for our Lagrangian are given by:
\begin{eqnarray}
  {\cal P}^{+0}&\!=\!&2\dot X^+\\
  {\cal P}^{-0}&\!=\!&2\big(\dot X^-+\i(\bar\theta\bar\gamma^-\dot\theta
    +\theta\bar\gamma^-\dot{\bar\theta})\big)
    -\big(X_a^2+4\bar\theta\bar\gamma^-\Pi\theta\big){\cal P}^{+0}\\
  {\cal P}^{I0}&\!=\!&2\cos(X^+)\dot X^I+\sin(X^+)X^I{\cal P}^{+0}\\
  {\cal J}^{+I0}&\!=\!&2\sin(X^+)\dot X^I-\cos(X^+)X^I{\cal P}^{+0}\\
  {\cal J}^{ij0}&\!=\!&2(X^{i}\dot X^j-X^j\dot X^i)-\i\bar\theta\bar\gamma^-
    \gamma^{ij}\theta{\cal P}^{+0}\\
  {\cal J}^{i^\prime j^\prime0}&\!=\!&2(X^{i^\prime}
    \dot X^{j^\prime}-X^{j^\prime}\dot X^{i^\prime})-\i\bar\theta\bar\gamma^-
    \gamma^{i^\prime j^\prime}\theta{\cal P}^{+0}\\
  {\cal Q}^{+0}&\!=\!&2\bar\gamma^-\e^{\i X^+\Pi}{\cal P}^{+0}\theta\\
  \bar{\cal Q}^{+0}&\!=\!&2\bar\gamma^-\e^{-\i X^+\Pi}{\cal P}^{+0}\bar\theta.
\end{eqnarray}
If we compare these currents to those obtained for the tensile case
\cite{Metsaev:2001bj}, we see that the only difference apart from the fact that
only one of the two world-sheet directions contribute, is the absence of terms
proportional to $\epsilon^{ab}$. These terms are connected to the WZW-term
in \eqref{eqn:ksymfixedL} which vanishes in the tensionless limit. This has,
however, no effect on the conserved charges generating the transformations. 
In light-cone gauge, these charges are: 
\begin{eqnarray}
  P^+&\!=\!&p^+\\
  P^I&\!=\!&\int \cos(X^+){\cal P}^I+\sin(X^+)X^Ip^+\\
  J^{+I}&\!=\!&\int \sin(X^+){\cal P}^I-\cos(X^+)X^Ip^+\\
  Q^+&\!=\!&\int 2p^+\bar\gamma^-\e^{\i X^+\Pi}\theta\\
  \bar Q^+&\!=\!&\int 2p^+\bar\gamma^-\e^{-\i X^+\Pi}\bar\theta.
\end{eqnarray}
Using the SUSY-algebra, we then calculate the dynamical currents and
charges ${\cal Q}^{-0}$ and $Q^-$, and obtain:
\begin{eqnarray}
  {\cal Q}^{-0}=2{\cal P}^I\bar\gamma^I\theta+2\i p^+X^I\bar\gamma^I\Pi\theta 
  \mbox{~~~and~~~}
  \bar{\cal Q}^{-0}=2{\cal P}^I\bar\gamma^I\bar\theta-2\i p^+
  X^I\bar\gamma^I\Pi\bar\theta .
\end{eqnarray}
Since
\begin{eqnarray}
  \del_0{\cal Q}^{-0}&=&2p^+\left(\ddot X^I+m^2 X^I\right)\bar\gamma^I\theta,
\end{eqnarray}
where we introduced the parameter $m$, we find that ${\cal Q}^{-0}$ is time
independent using \eqref{eoma}. The corresponding charge is then derived by
integration over $\sigma$, just as in the above cases.

\begingroup\raggedright\endgroup

\end{document}